\documentclass[final]{raa_twocolumn}
\usepackage{graphicx,times}
\usepackage{natbib}
\usepackage{algpseudocode}
\usepackage{algorithm}
\usepackage{amssymb,amsmath}
\bibpunct{(}{)}{;}{a}{}{,}

\usepackage{hyperref}
\hypersetup{pdftitle = The title of my PDF, pdfauthor = My name, pdfsubject= The subject, pdfkeywords = keyword1 keyword2 keyword3}
\hypersetup{colorlinks = true, linkcolor = green, anchorcolor = red, citecolor = blue, filecolor = red, pagecolor = red, urlcolor = red}

\newcommand{\erfc}{\textrm{erfc}}

\begin{document}
\title{PhotoNs-GPU:A GPU accelerated cosmological simulation code}

\setcounter{page}{1}

\author{Qiao Wang, Chen Meng}

\institute{
Key Laboratory for Computational Astrophysics, National Astronomical Observatories, Chinese Academy of Sciences, Beijing 100101, China\\
School of Astronomy and Space Science, University of Chinese Academy of Sciences, Beijing 100049, China ${qwang@nao.cas.cn}$\\
\vs \no
{\small Received 0000 xxxx 00; accepted 0000 xxxx 00}
}

\abstract{
We present a GPU-accelerated cosmological simulation code, PhotoNs-GPU, based on algorithm of Particle Mesh Fast Multipole Method (PM-FMM), and focus on the GPU utilization and optimization. A proper interpolated method for truncated gravity is introduced to speed up the special functions in kernels. We verify the GPU code in mixed precision and different levels of interpolated method on GPU. A run with single precision is roughly two times faster that double precision for current practical cosmological simulations. But it could induce a unbiased small noise in power spectrum. Comparing with the CPU version of PhotoNs and Gadget-2, the efficiency of new code is significantly improved.  Activated all the optimizations on the memory access, kernel functions and concurrency management, the peak performance of our test runs achieves 48\% of the theoretical speed and the average performance approaches to $\sim 35\%$ on GPU. 
\keywords{methods: numerical – cosmology: theory - large-scale structure of universe}
}

\authorrunning{Q. Wang \& C. Meng}  
\titlerunning{PhotoNs-GPU}
\maketitle

\section{Introduction}
High resolution N-body simulations are essential tools to understand the formation and evolution of dark matter from sub-halos to large scale structure of our Universe~\citep{2008MNRAS.391.1685S, 2012MNRAS.426.2046A, 2020Natur.585...39W}. 
During the past 50 years, the dramatic increase in the size of the simulation is not only because of the rapid progress on the supercomputer hardware, but also be owed to the rapid development of the N-body solvers~\citep{Ishiyama:2012:PAN:2388996.2389003, 2016NewA...42...49H, 2017NatAs...1E.143Y, 2017ComAC...4....2P, 2020arXiv200303931C, 2020arXiv200714720I}. 

One of popular N-body method is the Fast multiple method (FMM), it has an attracting feature with a time complexity O(N),  ideal for performing extreme large simulation ~\citep{1987JCoPh..73..325G, 1999JCoPh..155..468C,  2002JCoPh.179...27D,  2014ComAC...1....1D}. Indeed some extremely large simulations have been performed with the FMM N-body solver, for instance, \citet{2017ComAC...4....2P} completed a cosmological simulation with two trillion particles on the Piz Daint, GPU supercomputer, using the FMM code of PKDGRAV ~\citep{2001PhDT........21S}.  Recently, a hybrid method of Particle Mesh Fast Multipole Method (PM-FMM) is introduced to cosmological simulations. Analogue to the TreePM method~\citep{1995ApJS...98..355X, 2002JApA...23..185B, DUBINSKI2004111, 2005MNRAS.364.1105S, 2009PASJ...61.1319I, 2018RAA....18...62W}, it calculates the short-range tree method by FMM. \citet{Wang_2021} presented the details of the method, its advantage and accuracy, with the code PhotoNs-2. Note, \citet{2020arXiv201003567S} also provides an alternative implementation in Gadget-4 code.

In the era of heterogeneous supercomputer, it is important to develop  efficient N-body solvers based on these heterogeneous systems, instead of homogeneous ones~\citep{10.1145/224170.224400, 10.1093/pasj/56.3.521}. In recent years, some Branes-Hut Tree ~\citep{1986Natur.324..446B} and FMM N-body solvers  have been successfully developed on Graphics Processing Unit (GPU) platforms ~\citep{10.1145/1654059.1654123, Hamada2010190TA, GABUROV20101119, 2014hpcn.conf...54B,13976, 2011arXiv1110.2921Y, 2012arXiv1209.3516Y}. 

In this paper, we develop a GPU-accelerated PM-FMM code based on PhotoNs-2, especially on the NVidia architecture,  which is a referred to as PhotoNs-GPU.

This paper is organized as follows. In section \ref{sec:algorithm}, we briefly introduce the algorithm of PM-FMM method. The detail of GPU implementation is discussed in section \ref{sec:gpu}. Two test runs are presented on performance and accuracy of PhotoNs-GPU in section \ref{sec:simu}. Finally, some potential issues are discussed in Section~\ref{sec:dis}. 

\section{algorithm and Code}
\label{sec:algorithm}
First, we begin with a brief review of the PM-FMM method developed by \citet{Wang_2021}, in particularly the relevant parts for GPU acceleration in following sections. A Gaussian transition function is employed to split gravity into a smoothed long-range and short-range part, $\phi=\phi_{long}+\phi_{short}$. The $\phi_{long}$ is satisfied to Poisson's equation which is solved by a Particle-Mesh method based on the convolution of density field $\rho(x)$ with Green function of long-range gravitational force on a regular mesh. Thus, $\phi_{long}(x)= \mathcal{F}^{-1}[ (\hat{\rho}_k/ k^2) \exp(-k^2/4r_s^2)] $, where $\hat{\rho}_k= \mathcal{F} [\rho(x)]$ is the density field in Fourier space and the long-range potential is smoothed by a Gaussian function with a split radius $r_s = 1.2~\Delta_{g}$~($\Delta_{g}\equiv$ BOXSIZE~/~N$_{\bf PM}$).

The short-range gravity is computed by a truncated FMM which dominates the most amount of computation and is accelerated on GPU in this work. Similar to a conventional FMM, all particles in a computing domain are organized into a tree structure and the finest tree cells (or tree nodes) point to continuous particle packs (or leaves). An Orthogonal Recursive Bisection (ORB) tree is employed and the particles belonged to a cell are equally assigned into two offspring cells till to leaves. The maximum particle number allowed in a leaf is constrained by the parameter of `MAXLEAF'. 

In FMM, gravity of a particle is not directly accumulated by tree nodes or particles. Instead, the computation is based on particle packs or leaves. First, the information of particle distribution in leaves is transferred into the multipole of those leaves by operator \textbf{P2M} (Particle to Multipole). Thus the multipole of lower nodes built up the higher one, \textbf{M2M} (Multipole to Multipole). This process is called PASS-UP. The interaction between multipoles is computed by operator \textbf{M2L} (Multipole to Local) and the local multipole is passed down to the lower nodes by  \textbf{L2L} (Local to Local). Finally, gravity of a particle in local leaf is determined by the local multipole of gravitational potential \textbf{L2P} (Local to Particle) and direct interaction from particles in neighborhoods \textbf{P2P} (Particle to Particle).

All operators are straightforwardly derived from multipole expansion of gravitational interaction and M2L is specified by equation
\begin{equation}
{\mathcal L}_{\mathbf n}({\mathbf z}_B) = \sum_{|{\mathbf m}|=0}^{p-|{\mathbf n}|}{\mathcal M}_{\mathbf m}({\mathbf z}_A){\mathcal D}_{{\mathbf n}+{\mathbf m}}({\mathbf z}_B-{\mathbf z}_A),
\notag
\label{eq:m2l}
\end{equation}
where ${\mathcal D}_{\mathbf n} \equiv \nabla^{\mathbf n} \psi(r) =f_{(n)} \bar{{\mathbf r}}^{\mathbf n} $ is a {\it traceless} operator. $\bar{{\mathbf r}}^{\mathbf n}$ is a  displacement tensor and the prefactors of 
\begin{equation}
f^{\rm inv}_{(n)}(r)= (-1)^{n}\frac{(2n-1)!!}{r^{2n+1}},
\notag
\label{eq:pref}
\end{equation}
for inverse-square law. But for truncated short-range gravity, it becomes complicated. One can compute the prefactor $f_{(n)}$ of any order $n$ by equation of
\begin{equation}
\begin{aligned}
(-1)^{n}r_s^{2n+1}f_{(n)}(x) = \frac{(2n-1)!!}{2^{{2n+1}}} \frac{ \erfc \left( x\right) } {x^{2n+1}}\\
+ \sum_{q=1}^{n}  \frac{2^{-q-n}(2n-1)!!}{(2n-2q+1)!!}\frac{e^{ -x^2 } }{\sqrt{\pi}x^{2q}},
\end{aligned}
\notag
\end{equation}
where ${x \equiv {r}/{2r_s}}$. We numerically truncate the function at cutoff radius $\sim 6~\Delta_g$, following a traditional Gaussian splitting approach. Meanwhile, the split function for P2P direction summation of truncation gravity is computed by
\begin{equation}
{\mathbf F_{s}}({\mathbf r}) = -\frac{\mathbf r}{r^3} T(r;r_s),
\notag
\end{equation}
where the truncation function reads
\begin{equation}
\label{eq:tr}
T \equiv \erfc \left( \frac{r}{2r_s} \right) + \frac{r}{r_s\sqrt{\pi}} \exp \left( -\frac{r^2}{4r_s^2} \right).
\end{equation}
We only summarize the relevant algorithms and formulas used for GPU acceleration in this section, and we refer the readers for more technical details to \citet{Wang_2021}.

\section{GPU implementation}
\label{sec:gpu}
In practise, the short-range gravity in PhotoNs-2 is estimated by walking the multipole tree to determine whether operators M2L or P2P between two multipole nodes/leaves need to be considered. We use two interaction lists to record the interaction pair of M2L and P2P by one traversal, respectively. Meanwhile, we identify a M2l or P2P pair as a {\bf task} so that the interaction lists naturally become  task queues. Usually, the lists are long enough to guarantee the concurrency on multi-core CPUs or GPUs. Two task lists are both dealt with on CPU in PhotoNs-2. In the GPU version, we push the P2P task on GPU, because of the dominant computation amount of P2P over others. 

\subsection{Memory layout}
The information of each particle is stored in a predefined data structure. It is suitable to communicate between computing domains, but does not match the coalesced memory access mechanism of GPU when loading data. Therefore we transfer an Array of Structure (AoS) of particles into Structure of Array (SoA) of particle components, e.g., position, velocity, acceleration etc. Since the particles are already reordered in the process of tree building, such a transformation is natural and widely used in many N-body problems with CUDA~\citep{Nyland2009}. The transformation executes on host memory, then the arrays of position is sent to GPU. 

\begin{figure*}[htbp]
\centering
\includegraphics[width=1.\linewidth,angle=0]{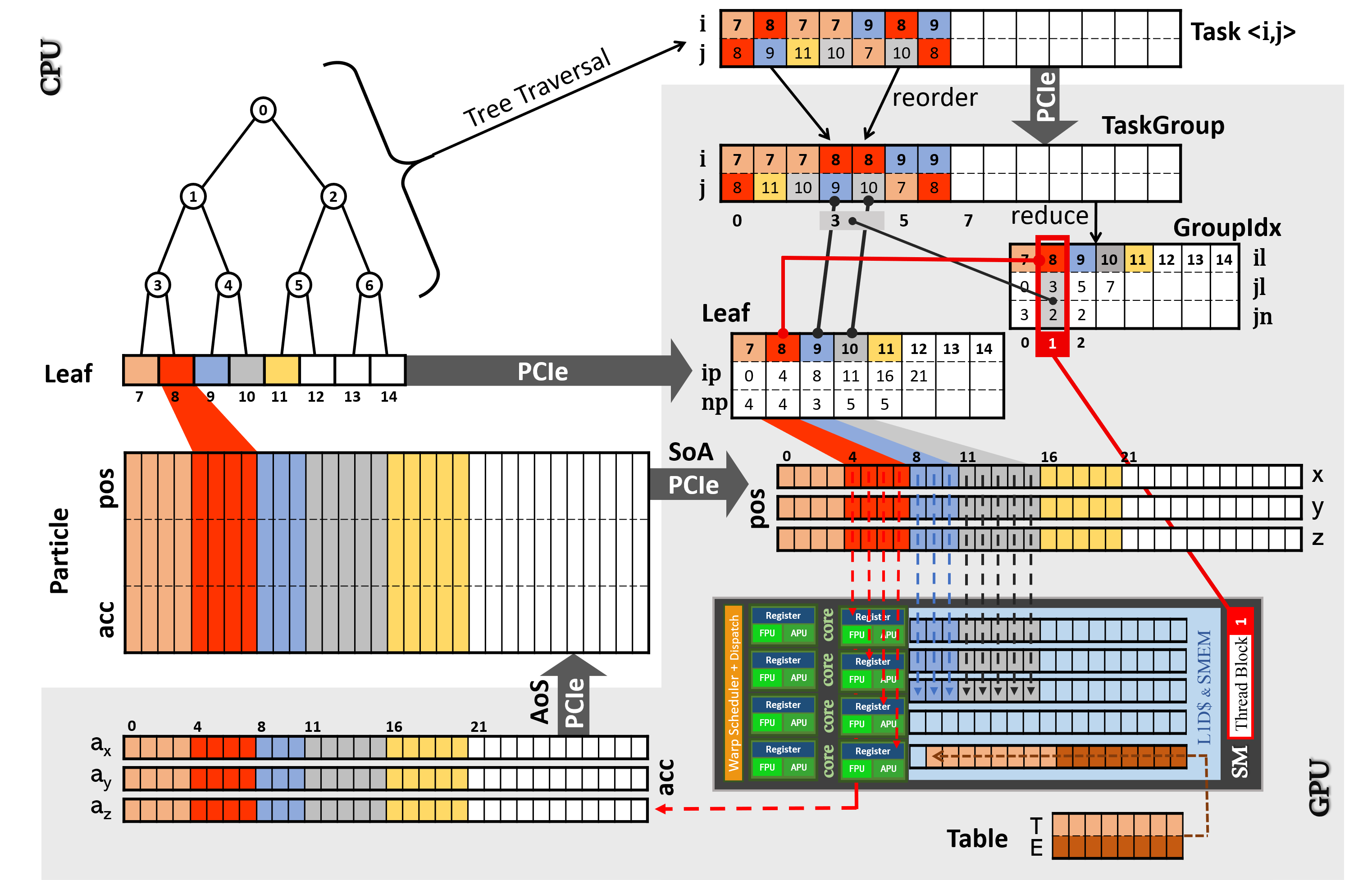}
\caption{Memory layout and relation on CPU and GPU (shadow zone). The dark gray box denotes a SM unit. Arrows indicates the flow of data.}
\label{fig:mem}
\end{figure*}

Fig~\ref{fig:mem} is schematic diagram of data arrays in the host memory and GPUs. There are three main data arrays for particle, leaf and tasks. Each particle structure contains position, velocity and acceleration, etc. We abstract the position information of x, y, z (Pos) into individual arrays (SoA) for further optimal GPU L2 cache performance. The task list (Task) of P2P interaction is an array of pair $<i,j>$ which indicates  attractive force of particles in $i$-leaf induced by $j$-leaf.

They are sent to the GPU global memory denoted by the gray shadow zone. One {\bf Streaming Multiprocessor} (SM) is shown with a dark gray box, which contains a chip of {\bf level-1 data cache} (L1D\$) plus {\bf shared memory} (SMEM) and a group of sub-cores with registers and Float Point Units, etc. We allocate two main arrays to store {\bf Particle} and {\bf Leaf}. Each leaf records starting index of first particle and particle number in the leaf.

\subsection{GPU kernel}
Arrays of Task list, Leaf and Pos are transferred to the global memory of GPU via PCIe port. The item with the same index in {\bf Task} list is not continuous in memory, which dramatically slows down the speed of data accesses into CUDA cores. To fix this problem, we reorder Task list with respect to $i$ and rename this list as {\bf TaskGroup}.  A CUDA library {\tt thrust} is utilized for this reordering. Meanwhile the index array of Task Group ({\bf GroupIdx}) is produced to record the starting points and length of tasks for every $i$-leaf in the task list.

According to the scheduling mechanism of GPU, we deal with all interactions in one task group with a thread block, which corresponds to a SM unit~\citep{10.1145/1654059.1654123}. Since every task group contains only one $i$-leaf and all particles in $i$-leaf are continuously stored, each thread responds to one particle in $i$-leaf and the particles in $j$-leaf are prefetched into Shared memory for interaction calculation.

For instance, we assume that the gravity of leaf 8 is induced by leaf 9 and 10 in Fig.~\ref{fig:mem}. As the 2nd column in `GroupIdx', Thread block 1 is correspondent to it. Four particles in leaf 8 is sent into the register of sub-cores and particle position in leaf 9 and 10 are sent to shared memory of SM. {\bf Core} function returns the acceleration acc$\_$i of leaf 8 (see implementation of Core function in Alg.~\ref{alg:core}).  After all calculation of gravity is done, array of acceleration is transferred back to host memory.

\begin{algorithm}[H]
\caption{P2P Kernel\label{alg:kernel}}
\begin{algorithmic}
\State \_\_shared\_\_ {\bf sh\_pos\_j}
\For{$b \leftarrow$ blockIdx.x \textbf{to} $len$(GroupIdx) } 
    \State i$\_$leaf = GroupIdx.il[b]
    \State $t$ = threadIdx.x
\If {$t$ $\textless$ Leaf[i\_leaf].np}
    \State load {\bf pos}[$t$ + Leaf[i$\_$leaf].ip] to register {\bf pos\_i}
    \State reset register acc\_i with 0
\EndIf
    \State j\_start =  GroupIdx.jl[b]
    \State j\_end = GroupIdx.jl[b] + GroupIdx.jn[b]  
\For{$j \leftarrow$ j\_start \textbf{to} j\_end } 
    \State  j\_leaf = TaskGroup.j[$j$]
    \State \_\_syncthreads\_\_
    \If {$t$ $\textless$ Leaf[j\_leaf].np}
    \State preload {\bf pos}[$t$+Leaf[j\_leaf].ip] 
    \State ~~to SMEM {\bf sh\_pos\_j}
    \EndIf
    \State \_\_syncthreads\_\_
    \State nj = Leaf[j\_leaf].np    
    \State acc\_i $+=$ Core({\bf pos\_i}, {\bf sh\_pos\_j}; nj) 
    \State  \Comment{goto Algorithm \ref{alg:core}}
\EndFor 
\State accumulate register acc\_i to global memory \bf{acc}
    \State $b$ += gridDim.x
\EndFor \Comment{block level parallelism}
\end{algorithmic}
\end{algorithm}

\subsection{Interpolation function}
In previous section, we present the kernel and memory layout on GPU. In function of {Core}, two special functions are necessary in this kind of splitting method. Compared with the computation of pair-wise inverse-square gravity, they seriously decrease the efficiency of P2P kernel. The Gaussian shaped prefactor contains an exponential function ({\tt exp}) and an error function ({\tt erf}). Unlike function {\tt rsqrt} estimated by an `intrinsic' implementation, {\tt exp} could consume about 5.5 times the operations of {\tt add/mul} in {\bf fast math} library on a NVidia GPU. We catch the effective number of floating-point operations (flops) by the tool of {\bf nvprof} and find that {\tt erf} needs to implicitly call functions of {\tt rcp}, {\tt exp} and several extra operations. This estimation is also consistent with the measurement of \citet{2019arXiv190508778A}.

In order to speed up the evaluation of special functions, we combine those two functions into a table for the interpolation algorithm. The truncation function (Eq.~\ref{eq:tr}) can be transformed to an integral form, 
\begin{equation}
T(x\equiv \frac{r}{2 r_s}) =\frac{4}{\sqrt{\pi}} \int_x^{\infty} u^2 e^{-u^2} du,
\notag
\end{equation}
by using partial integration. Obviously, its Taylor's series reads
\begin{equation}
\label{eq:int}
\begin{aligned}
T(x) &= T_i & :T_{[0]}\\
&+  x_i^2 E_i\epsilon  & :T_{[1]}\\
&- x_i (x_i^2 - 1)E_i \epsilon ^2 & :T_{[2]} \\
&+ (\frac{2}{3} x_i^4 - \frac{5}{3} x_i^2+\frac{1}{3}) E_i\epsilon^3 &  :T_{[3]}\\
&- x_i (\frac{ 1}{3}x_i^4 - \frac{3}{2} x_i^2 +1  ) E_i\epsilon ^4 &  :T_{[4]}\\
&+ \mathcal{O}(\epsilon^5), & ~
\end{aligned}
\end{equation}
where $\epsilon = x - x_i$ , $E_i = - 4{\exp}({-x_i^2} )/ {\sqrt{\pi}} $ and  $ T_i =\erfc \left( x_i \right) - x_i E_i / 2$ at the $i$-th gird of table in the range of [0,3]. In this case, we set grid number of $E_i$ and $T_i$ to 512, considering the suitable capacity of shared memory (light and heavy brown arrays in Fig.~\ref{fig:mem}).

\begin{figure}[htbp]
\centering
\includegraphics[width=0.9\linewidth]{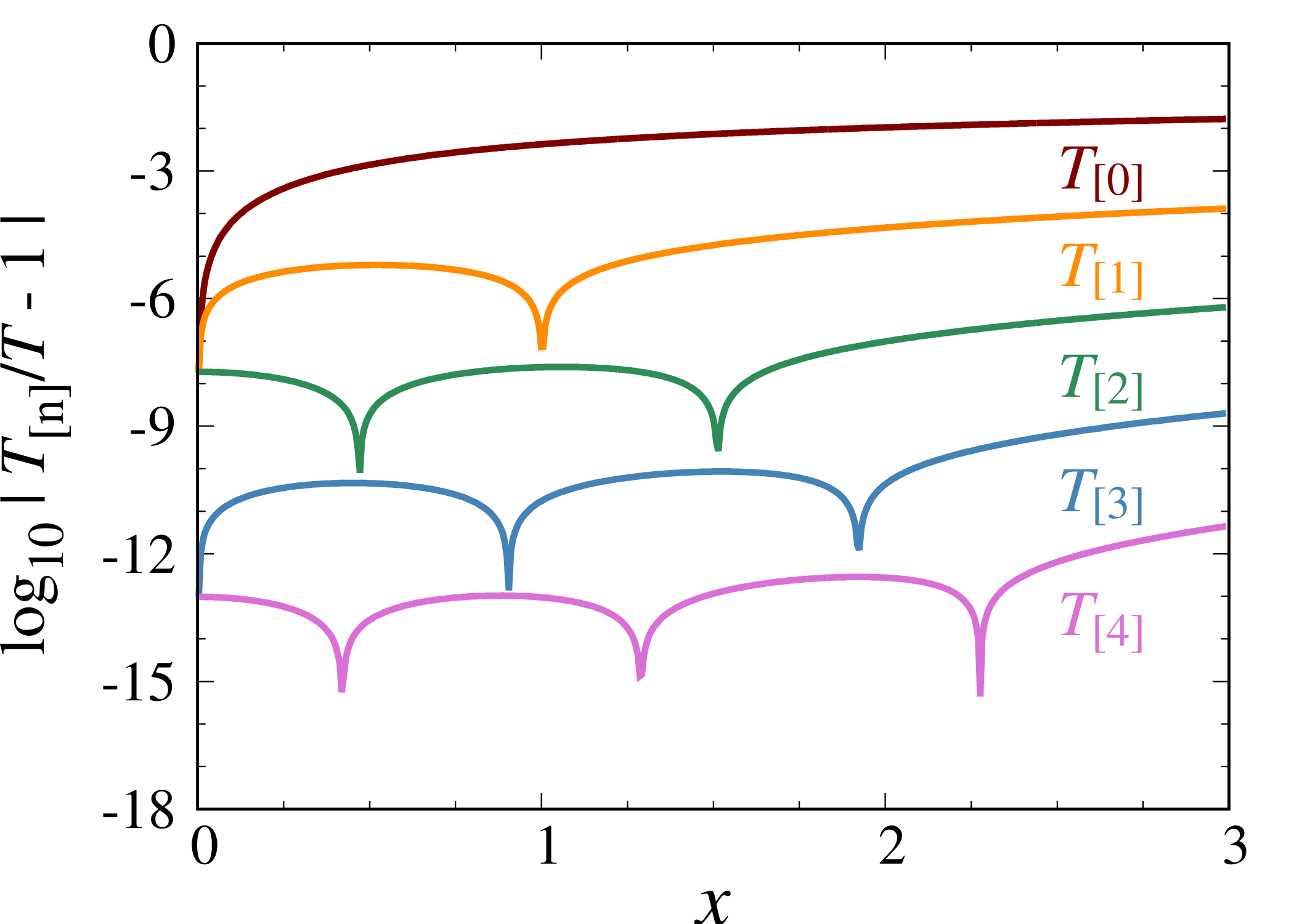}
\caption{From top to bottom, The curves denote the maximum envelop of relative error with respect to exact splitting function at the interpolation level of $T_{[0]}$, $T_{[1]}$, $T_{[2]}$, $T_{[3]}$ and $T_{[4]}$, respectively.  }
\label{fig:err}
\end{figure}

We measure the precision of interpolation method to link accuracy with interpolation level. Fig.~\ref{fig:err} shows that the maximum  value of relative errors at different levels. We show the absolute value so that the transition of positive and negative value occurs at the position of the cusps of curves. According to the measurement result, $T_{[2]}$ is adequate for the single-precision calculation and  $T_{[4]}$ is for double-precision. In the next section, we run some practical simulations to observe the overall interpolation effect. 

Eq.~\ref{eq:int} contains thirty nine floating point operations and $loads$ two float number from shared memory. The floating point operations can be reduced to fourteen by variable reuse and Fused Multiply-Add (FMA). Combining with other optimizations, the amount of flops can be suppressed by $\sim 60\%$ in total. The modified `Core' is presented in Alg. ~\ref{alg:core}. 
\begin{algorithm}[H]
\caption{Core function\label{alg:core}}
\begin{algorithmic}
\Function{Core}{pos\_i, sh\_pos\_j; nj}
\For{$j \leftarrow$ 0 \textbf{to} nj } 
\State pos$\_$j = sh\_pos\_j[$j$] \Comment{read from SMEM}
\State dx = pos$\_$j.x - pos$\_$i.x
\State dy = pos$\_$j.y - pos$\_$i.y
\State dz = pos$\_$j.z - pos$\_$i.z
\State r2 = dx*dx + dy*dy + dz*dz
\State determine table index $d$ and $\epsilon$ in Eq.~\ref{eq:int}
\State load $T_d$
\State load $E_d$
\State idr = rsqrt(r2)
\State pref ~~= $T_{[0]}(d,\epsilon)$
\State pref += $T_{[1]}(d,\epsilon)$
\State pref += $T_{[2]}(d,\epsilon)$ \Comment{single precision}
\State pref += $T_{[3]}(d,\epsilon)$
\State pref += $T_{[4]}(d,\epsilon)$ \Comment{double precision}
\State grav = GM*idr*idr*idr*pref
\State ax = grav*dx
\State ay = grav*dy
\State az = grav*dz
\EndFor
\State return acc\_i = (ax, ay, az)
\EndFunction         \Comment{thread level parallelism}
\end{algorithmic}
\end{algorithm}

\section{test run}
\label{sec:simu}
We modify the CPU-based PM-FMM code to a GPU-accelerated version that is referred to as {\bf PhotoNs-GPU}. For testing the precision and efficiency of our code, we carry out two groups of cosmological simulations on a 1.5 TB memory GPU server with thirty-two cores of Intel Xeon 5218 CPU (2.3 GHz) and two NVidia Tesla 32GB V100S GPUs. The theoretic single precision (SP) performance of the machine is about 32.7 TFlops and the PCIe bandwidth is about 10 GB/s for each GPU.

\subsection{Precision check}

We test the accuracy of our code by comparing simulations run with {\bf Gadget-2} and PhotoNs-GPU. We employ $256^3$ particle in 1 $h^{-1}$Gpc simulation box. The initial condition at $z_i=99$ is generated by {\bf 2LPTic} ~\citep{2006MNRAS.373..369C}, which is evolved for four simulations by Gadget-2 and PhotoNs-GPU, respectively.

\begin{figure}[htbp]
\centering
\includegraphics[width=0.9\linewidth]{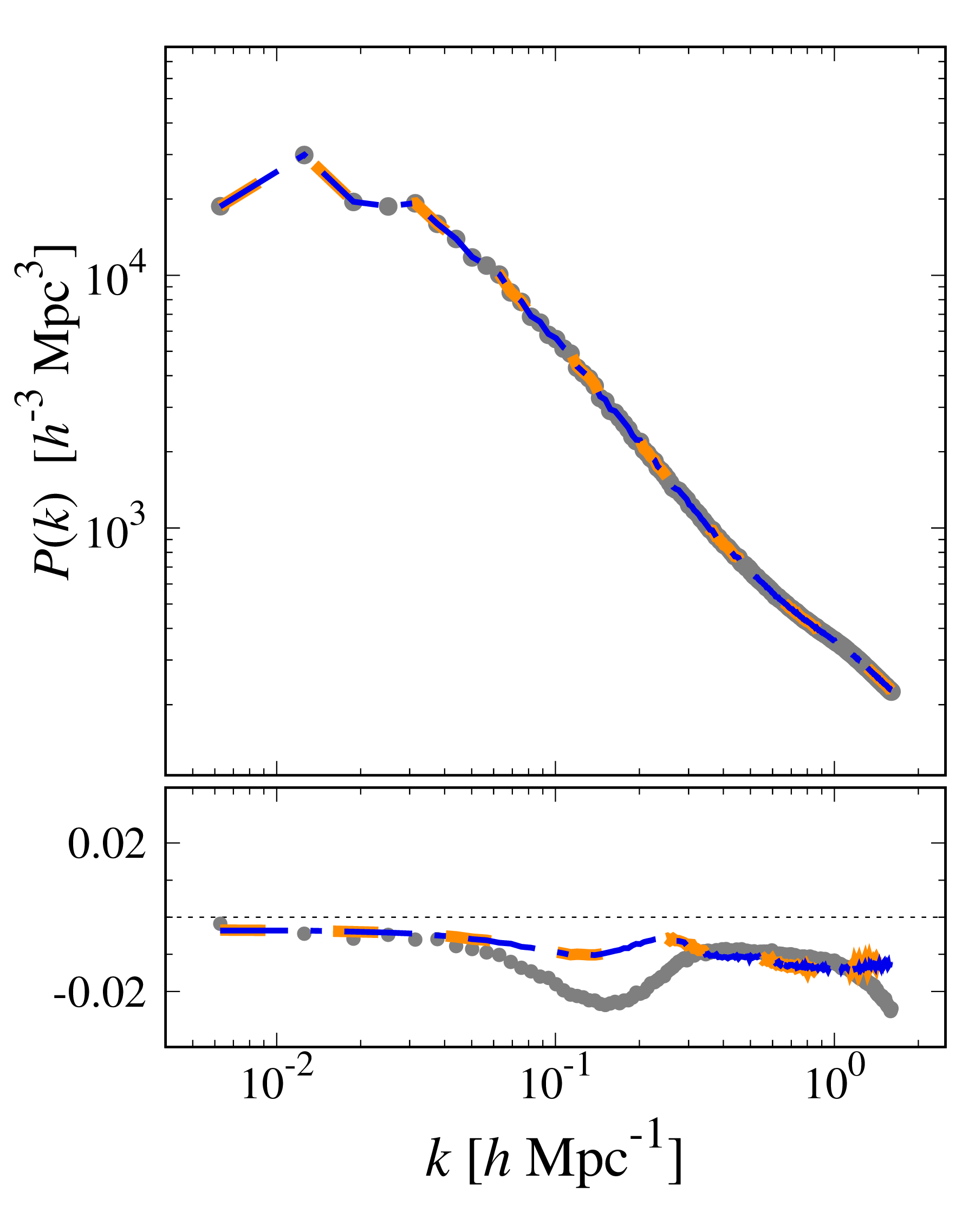} 
\caption{The comparison of power spectrum at z=0. The gray points denote Gadget-2 result, the heavy orange dashed line denotes SP run and the blue dash-dotted line denotes DP run by PhotoNs-GPU. In the bottom panel, the residuals of power spectrum show that the DP and SP runs are both consistent with the fiducial one.}
\label{fig:psc}
\end{figure}

We use 16 MPI processes to carry out the simulation to the present $z=0$ with N$_{\bf PM}=256$ and $100~h^{-1}$kpc softening length. Gadget-2 executes single step by 59 seconds of wall clock time and about 11,000 seconds in total. The power spectrum at $z=0$ is indicated with gray points in Fig.~\ref{fig:psc}. Based on the same parameters and settings, we run two different simulations to check the influence of floating point precision on GPU. The calculation on GPU with single-precision (SP) is indicated with Golden dashed curve and double-precision (DP) is blue dash-dotted curve. In the same condition, PhotoNs-GPU executes single step by $\sim 1.3$ seconds of wall clock time and about 1,000 seconds in total, using 16 MPI processes and two V100 GPUs.

The forth simulation is carried out as a fiducial one, which adopt direct P2P method to replace the Tree or FMM in order to exactly compute short-range gravity. Meanwhile we increase time steps by roughly three times. The bottom panel of Fig.~\ref{fig:psc} is the residual of power spectrum with respect to the fiducial simulation. The contrast of power spectrum is smaller than 1$\%$. Fig~\ref{fig:psc} shows that the SP case is well consistent with DP one, but more noisy at small scale ($k \sim 1$). Since PhotoNs-GPU employs a larger cutoff radius and more P2P contribution for gravity solver, our accuracy is slightly better than Gadget-2 in power spectrum statistics at the same settings.

\subsection{Planck's cosmology}
According to the recent observation of the best-fit 'plik' cosmological parameters from Planck 2018~\citep{2020A&A...641A...6P}: $\Omega_{\mathrm \Lambda}$=0.684, $\Omega_{\mathrm c}$= 0.265,  $\Omega_{\mathrm b}$ = 0.0494, $H_0$= 67.32 $km~s^{-1}$Mpc$^{-1}$, $\sigma_{8}$=0.812 and $n_s$ = 0.966, we employ $512^3$ particles in 100 $h^{-1}$Mpc simulation box. Since the test in the previous section indicates that a mixed-precision run is also consistent with the fiducial results in DP, we set all variable as SP to reduce the memory occupation and amount of communications and to utilize more Floating points units. The execution time is about five hours of wall clocks from redshift $z=99$ to $z=0$, using sixteen MPI processes and two Tesla V100S GPUs.

\begin{figure*}[htbp]
\centering
\includegraphics[width=0.49\linewidth]{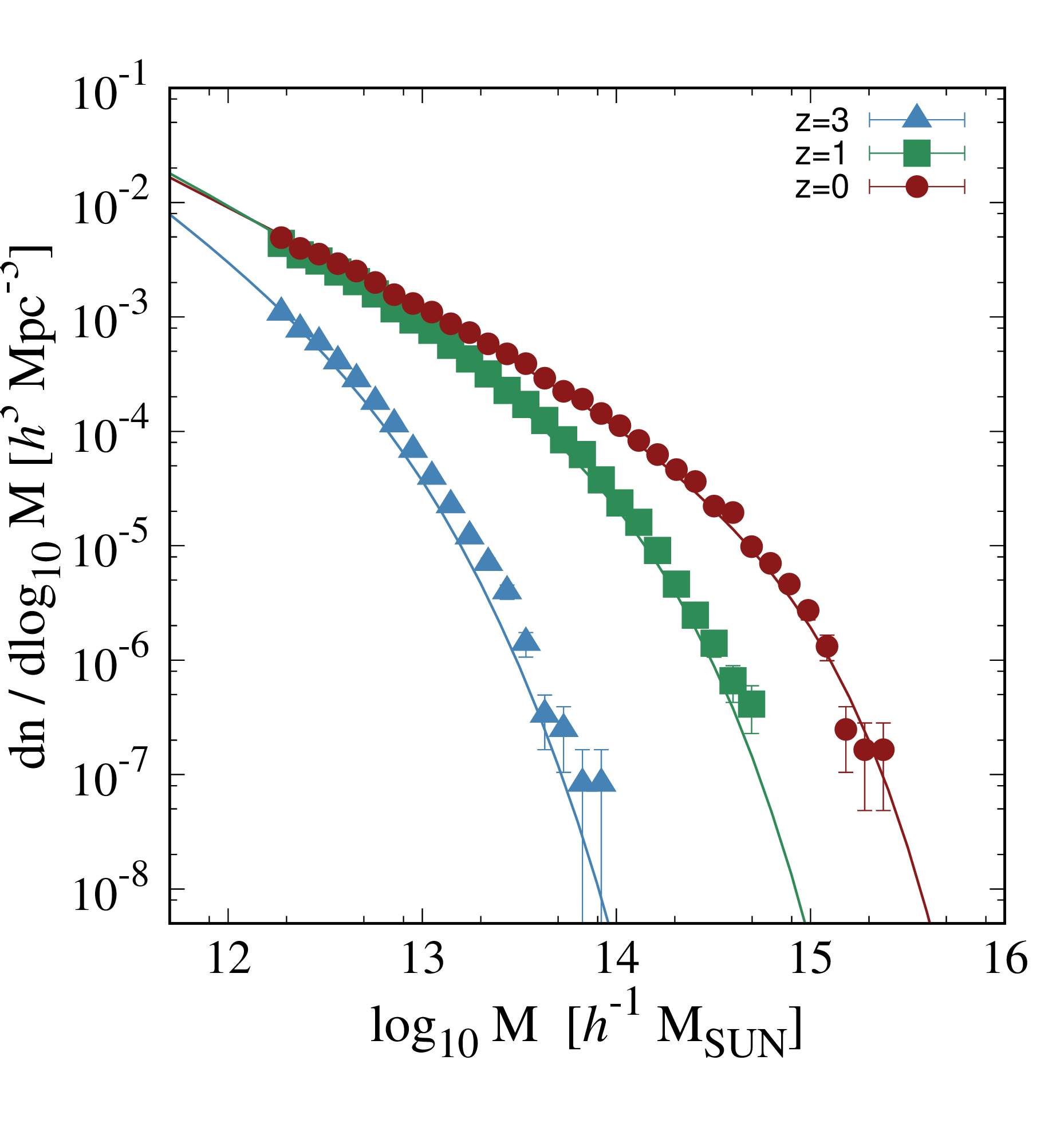}
\includegraphics[width=0.49\linewidth]{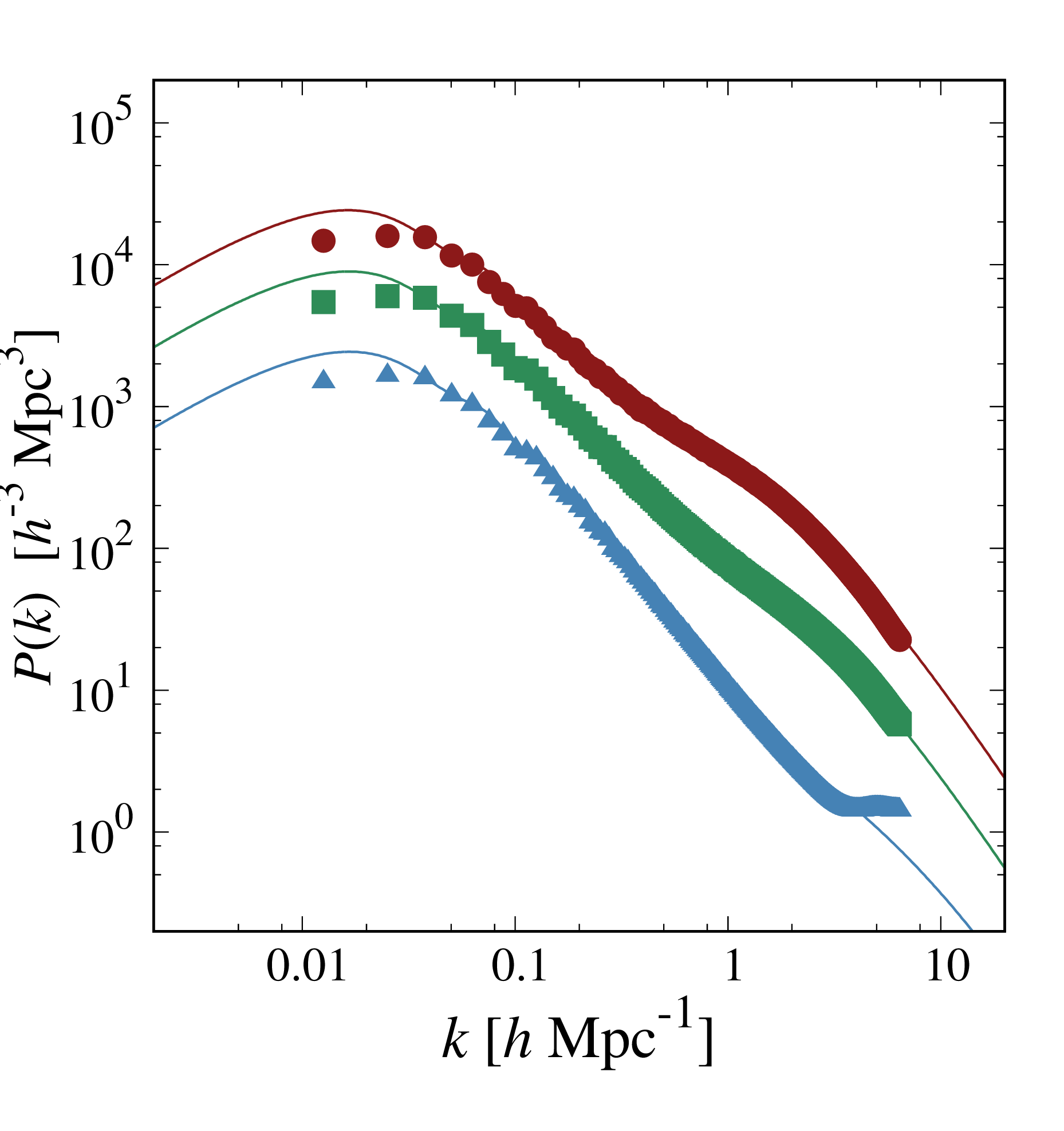}
\caption{The comparison of mass function (left panel) and power spectrum (right panel) of test run with theoretical curves. The (red) points, (green) squares and (blue) triangles denote the simulation results at $z=0, 1, 3$, respectively.}
\label{fig:mfps}
\end{figure*}

The initial condition is constructed by a built-in IC module, following the Zel'dovich approximation method\citep{2015ascl.soft02003S} to effectively match our domain decomposition. The built-in module is based on the subroutine of 2LPTic and the notations are following the appendix A of \citet{2006MNRAS.373..369C}. The input initial power spectrum is calculated by CAMB \citep{Lewis:1999bs}.

Dark haloes mass functions at redshift $z=0,1,3$ are shown in the left panel of Fig.~\ref{fig:mfps} and power spectra in the right panel. The mass function is measured by an on-the-fly Friend-of-Friend (FoF) halos finder and the theoretical curves are produced by \citet{2007MNRAS.374....2R}. The non-linear power spectrum is predicted by the HALOFIT \citep{2012ApJ...761..152T, 2016MNRAS.459.1468M}. The deviation of power spectrum of  $z=3$ (triangles) occurs at scale of $k\sim 5 $. It could be due to the lack of sampling. Higher resolution test shows that that deviation disappears and the simulation result is well consistent with the predictions.  

\section{discussion}
\label{sec:dis}

We present a GPU implementation of the PM-FMM algorithm by using an interpolated improvement for truncated pairwise interactions, then we tune the data structure to speed up GPU memory accesses and adjust the order of instructions in the kernel. A SP level test shows that SP runs can catch adequate accuracy on the statistics of large scale structure.

Briefly, after memory arrangement, interpolation and instruction fine tuning, the efficiency of code is significantly improved. In the case of the  $512^3$ simulations with Planck's cosmology , the number of particle-particle interactions per step averaged in the first ten steps is $\sim 1.74 \times 10^{12}$. The mean wall-clock time per step is 7.62 seconds, in which the kernel time is 5.5 seconds. Thus, the average performance is 11.65 TFlops and the peak performance of the kernel is 16.134 TFlops. Here, we use the operation count of 51 per interaction. On our 32.7 TFlops server, the measured efficiency of averaged and peak performance reaches 35\% and 48\% on GPU, respectively.

Although the optimizations and tests in this work are carried out on a small server, PhotoNs-GPU is potentially powerful for massive parallelism on supercomputers. Therefore we add a built-in IC generator and an on-the-fly halo finder~\citep{2020RAA....20...46S} into the code to save data storage. More on-the-fly data processes will be included into the code in the future. 

We only focus on the P2P operator of the algorithm on GPUs in this work. We measure its performance and verify the feasibility for cosmological simulations. But there still needs more considerations and improvements on the other functional modules. For instance, M2L operator becomes more important for small box simulations or in isolated boundary conditions. Similarly, FFT in PM method could become dominant in extremely large simulations. Further optimizations and considerations about them will be done in our future works. 

\normalem
\begin{acknowledgements}
We acknowledge the support from National SKA Program of China (Grant No. 2020SKA0110401), National Natural Science Foundation of China (Grant No. 12033008) and K.C.Wong Education Foundation. WQ thanks the useful discussions with J. Makino, M. Iwasawa, and L. Gao.
\end{acknowledgements}

\bibliographystyle{raa}

\clearpage
\end{document}